\documentclass[twocolumn,preprintnumbers,superscriptaddress,nofootinbib,aps,prd,floatfix]{revtex4-2}
\usepackage{enumerate}
\usepackage{amsmath}
\usepackage{amssymb}
\usepackage{graphicx}
\usepackage{hyperref}
\hypersetup{colorlinks=true, citecolor=blue, urlcolor=blue, linkcolor=blue}
\usepackage{subfigure,orcidlink}

\begin{document}
\title{Weak boson probes of Higgs unitarity restoration at 10 TeV parton colliders} 
\allowdisplaybreaks
\flushbottom
\begin{abstract}
Higgs coupling deviations, at levels accessible to the high-luminosity LHC, can imply a phenomenological no-lose theorem for the next generation of collider facilities. Correlating Higgs coupling deviations from the SM expectation in the gauge boson sector with high-scale unitarity requirements, we estimate and compare the sensitivity that can be expected at a future hadron collider (operating at 100 TeV centre-of-mass energy) and a 10 TeV muon collider. Both muon and hadron colliders offer discovery potential for mass scales up to ${\cal{O}}(6~\text{TeV})$ where unitarity violation induced by (sub)percent Higgs coupling modifications is mended. We comment on how an intermediate precision FCC-ee programme can corroborate such deviations.
\end{abstract}
\author{Christoph~Englert\orcidlink{0000-0003-2201-0667}}
\affiliation{Department of Physics \& Astronomy, University of Manchester, Manchester M13 9PL, United Kingdom\\[0.1cm]}
\author{Wrishik~Naskar\orcidlink{0000-0002-4357-8991}}
\affiliation{Deutsches Elektronen-Synchrotron DESY, Notkestra\ss e 85, 22607 Hamburg, Germany\\[0.1cm]}
\author{Andrew D. Pilkington\orcidlink{0000-0001-8007-0778}}
\affiliation{Department of Physics \& Astronomy, University of Manchester, Manchester M13 9PL, United Kingdom\\[0.1cm]}
\author{Michael Spannowsky\orcidlink{0000-0002-8362-0576}}
\affiliation{Institute for Particle Physics Phenomenology, Department of Physics, Durham University,\\Durham DH1 3LE, United Kingdom\\[0.1cm]}
\preprint{DESY-26-006}
\preprint{IPPP/25/23}
\maketitle
\allowdisplaybreaks
\section{Introduction}
\label{sec:intro}
The Higgs boson plays a clear and prominent role in the landscape of electroweak symmetry breaking. As the residual dynamical scalar degree of freedom after spontaneous electroweak symmetry breaking, the Higgs plays a crucial part in the perturbative unitarisation of massive particle scattering processes in the Standard Model (SM)~\cite{Lee:1977yc,Lee:1977eg,Chanowitz:1978mv,Chanowitz:1985hj}. This feature defined a no-lose theorem for the Large Hadron Collider (LHC): either observe a Higgs boson below the TeV scale or witness signatures of strong electroweak symmetry breaking as the longitudinal gauge boson polarisations become strongly coupled. Informed by precision analysis during the Large Electron Positron (LEP) collider era, Nature appears to favour a perturbative TeV scale, as evidenced by the observation of $m_H\simeq 125~\text{GeV}$ and couplings consistent with the SM expectation. Furthermore, as the SM is a complete renormalisable theory, no equally strong no-lose theorem can be formulated post-discovery.

The LHC is expected to probe Higgs couplings at the 2\% level using the high-luminosity (HL) dataset~\cite{deBlas:2019rxi,ATLAS:2025eii}. A possible experimental outcome approaching the end of the HL phase could be a slight tension between Higgs coupling measurements and the SM expectation. Such tension in the Higgs boson couplings to massive gauge bosons would reintroduce perturbative unitarisation as a guiding principle for new physics discovery. Specifically, by fixing the mass scale $m_{\text{NP}}$ relevant for unitarity restoration, perturbative probability conservation fixes the size of the new physics couplings, and together, these determine the new physics discovery potential. It is important to highlight that even when the Higgs boson appears to be perturbatively coupled, this still includes theories of strong electroweak symmetry breaking.

If this scenario is realised, it is natural to ask ``Which future collider concept will best prepare us for clarifying the origin of the deviation?''. On the one hand, a precision electron-positron machine operating at a moderate centre-of-mass energy to maximise $Z$-associated Higgs production will serve the purpose of cross-validating the HL-LHC outcome, increasing the statistical significance of the coupling deviation. On the other hand, exploration machines, operating at a partonic centre-of-mass energy of $\sqrt{\hat{s}}=10~\text{TeV}$, offer direct sensitivity to the new physics scales whilst maintaining sensitivity to the 125 GeV Higgs boson interactions. Two main contenders fall into this second category: a future hadron-hadron collider such as the FCC-hh, providing proton-proton collisions at a $\sqrt{s}= 100~\text{TeV}$~\cite{FCC:2018vvp}, and a muon collider operating at $\sqrt{s}=10~\text{TeV}$~\cite{Casarsa:2024kct} (see also~\cite{Liu:2023jta}). 

\newcommand{\mucol}{$\mu$ coll}
It is the purpose of this article to provide a comparison between the FCC-hh and a 10 TeV muon collider (\mucol) for the direct discovery of gaugephilic resonances that play the role of unitarity restoration for a given Higgs-coupling modification away from the SM. We will focus exclusively on weak boson fusion (WBF), which is the key process to observe such resonances and facilitates a direct comparison between the two concepts. Associated production of gaugephilic resonances provides another avenue, which is predominantly relevant at the FCC-hh, as it accesses a wider range of relevant partonic initial state energies. We will not consider this process in this work, which is also suppressed for the WBF selection criteria. Within reasonable assumptions, (spoiler alert) we will observe relatively similar sensitivity expectations. This reinforces a modern incarnation of the no-lose theorem: if Higgs-coupling deviations persist, multi-TeV resonances are expected to become accessible at next-generation facilities.

\begin{figure*}[!t]
\centering
{\includegraphics[width=0.6\textwidth]{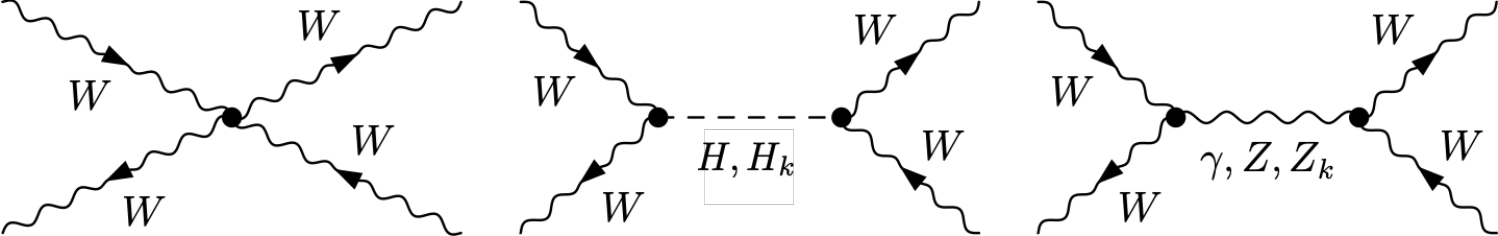}}
\caption{\label{fig:diags} Feynman diagram contributions to the $s$ channel of $WW\to WW$ scattering (suppressing $t$ topologies). Similar diagrams exist for $WZ$ scattering with $W$ resonances appearing as $s$ and $t$ channel contributions.}
\end{figure*}

This note is organised as follows: in Sec.~\ref{sec:coupd}, we review models that enable unitarity restoration as a function of Higgs couplings with minimal particle content that can evade LHC bounds if particles are heavy enough. Section~\ref{sec:sim} details our comparison, and Sec.~\ref{sec:res} provides numerical results for the discovery potential for the scenarios detailed in Sec.~\ref{sec:coupd}. There, we also clarify the change of sensitivity when the model's gaugephilic nature is relaxed, which can again be understood from a unitarity perspective (Sec.~\ref{sec:coupd}). In passing, we also comment on the impact of a $Z$ pole programme at a likely intermediate $e^+e^-$ machine. We provide conclusions in Sec.~\ref{sec:conc}.

\section{Higgs Couplings, Unitarity and Resonances}
\label{sec:coupd}
We focus first on the high-energy behaviour of WBF scattering. Even when the Higgs couplings are modified with respect to the SM expectation, the majority of WBF scattering does not reflect the naive longitudinal scattering $\sim s$ growth except in regions of phase space that are well modelled by the effective $W/Z$ approximation~\cite{Dawson:1984gx,Ruiz:2021tdt,Englert:2023uug}. That said, these regions exist, and unitarity violation is avoided through resonances in the spectrum: in the SM, the Higgs boson plays this crucial role, restoring high-energy unitarity above the Higgs threshold. Earlier work by Birkedal, Matchev and Perelstein provided coupling selection rules to ensure a good high-energy behaviour of WBF scattering via vectorial iso-triplet resonances~\cite{Birkedal:2004au}. These selection rules were particularly relevant for the so-called higgsless models and their holographic descriptions~\cite{Witten:1998qj,Klebanov:1999tb,Arkani-Hamed:2000ijo,Arkani-Hamed:2000ijo,Rattazzi:2000hs}, where the sum rules arise as completeness relations of the compactified theory~\cite{Csaki:2003zu,Csaki:2003dt,Csaki:2005vy,Cacciapaglia:2006mz}. Although Higgsless theories are now ruled out by the observation of the Higgs boson, compositeness is still a viable option for BSM physics. In such an instance, new resonances in addition to the Higgs boson appear above the 125 GeV mass scale~\cite{Pomarol:2012qf}, e.g., similar to the appearance of the $\rho$ mesons in QCD, which are roughly five times heavier than the pions. Such a situation can be phenomenologically parametrised as~\cite{Cacciapaglia:2006mz,Alboteanu:2008my,Englert:2015oga}
\begin{subequations}
\label{eq:unitr}
    \begin{alignat}{5}
      \label{ww1}
      g_{WWWW}&= g_{WW\gamma}^2+\sum_{k\geq 1} g_{WWZ_k}^2\,, \\
      \label{ww2}
      4m_W^2g_{WWWW} &= \sum_{k\geq 1} 3 m_{Z_k}^2 g_{WWZ_k}^2
     + \sum_{k\geq 1} g_{WWH_k}^2 \,,
\end{alignat}
for $WW$ scattering shown in Fig.~\ref{fig:diags}. $k$ runs over all resonances; the SM ones are identified as $H_1=H$, etc. For $WW\to ZZ$ (and $W^\pm Z\to W^\pm Z$) scattering, the sum rules are
 \begin{alignat}{5}
      \label{wz1}
      g_{WWZZ}& =\sum_{k\geq 1} g_{W_kWZ}^2\,,  \\
      \label{wz2}
      2(m_W^2+m_Z^2)g_{WWZZ} &= \nonumber \\& \hspace{-1cm} \sum_{k\geq 1} \left(3 m_{W_k}^2 - {(m_Z^2-m_W^2)^2\over
          m_{W_k}^2} \right) g_{W_kWZ}^2  \nonumber \\ &  + \sum_{k\geq 1} g_{WWH_k}g_{ZZH_k}   \,.
\end{alignat}
\end{subequations}
These sum rules cancel the $s^2$ (Eqs.~\eqref{ww1} and \eqref{wz1}) and~$s$ (Eqs.~\eqref{ww2} and \eqref{wz2}) high-energy behaviour of longitudinal gauge boson scattering above the relevant particle thresholds $i$. Mixed longitudinal-transverse scattering does not lead to new rules. As we will be focusing on WBF, we will assume, for the most part, that the extra vector states are fermiophobic. Significant couplings of $W_k, Z_k$ to SM fermions for $k>1$ motivate Drell-Yan production, which sets complementary constraints to the production modes discussed here (which are also not directly comparable between muon and pp colliders). We will also ignore couplings of type $Z_{k>1} Z H$. In the fermiophobic limit of the new gauge bosons, such contributions would change the total decay width; the latter is likely more dominated by the interactions with massive fermions, which we will comment on further below.\footnote{In holographic approaches, the effective couplings arise from overlap integrals along the compactified direction of the 5d gauge theory; the sum rules arise as orthonormality relations of the underlying Sturm-Liouville problem, e.g.~\cite{Csaki:2005vy}.}

It is instructive to see how these sum rules are obeyed in the SM. Firstly, due to gauge invariance, we have $g_{WWWW}=g_w^2$ (the $SU(2)_L$ gauge coupling, $g_w=e/s_W$\footnote{We denote $s_W,c_W$ as the sine and cosine of the Weinberg angle, respectively.}), and $g_{WWZ}=g_w c_W, g_{WW\gamma}=g_w s_W = e$. The sum rule for the $s^2$ behaviour is, therefore, satisfied by gauge invariance. Secondly, the $s$-divergent behaviour is mended in the SM as a result of electroweak symmetry breaking with couplings related to the vacuum expectation value, e.g., $g_{WWH}=g_w m_W$. Notably, longitudinal $ZZ$ scattering is not sensitive to the mechanism of electroweak symmetry breaking (except for mixing effects, see below).\footnote{Abelian masses can also be introduced semi-classically via a St\"uckelberg mechanism~\cite{Stueckelberg:1938hvi,Ruegg:2003ps} (at very different quantum properties compared to the SM weak sector).} 

Vector and scalar resonances exhaust the possibilities for ameliorating or curing harmful high-energy behaviour. Tensor resonances inevitably lead to pathological high-energy amplitudes~\cite{Alboteanu:2008my}, and we do not consider them further. These observations are semi-classical manifestations of the quantum behaviour of spontaneously broken quantum field theories~\cite{Cornwall:1973tb,Cornwall:1974km}. Theories with a well-behaved high-energy limit to all orders have no choice but to replace virtual longitudinal gauge boson contributions by derivatively coupled scalar degrees of freedom to restore the superficial degree of divergence counting. This sets tight constraints that shape tree-level approximations, which dominate phenomenological implications, e.g., as demonstrated by the SM itself. Any theory of unitarising iso-vectors is, therefore, only valid as an effective theory below some cut-off. This is obvious for the 5D theories~\cite{Cacciapaglia:2006mz} underpinning Eq.~\eqref{eq:unitr}; tensorial states further enhance the energy growth of amplitudes.

In this work, we will not refer to any particular underlying renormalisable or effective theory. Instead, the above relations enable a consistent treatment of the high-energy environment that is explored at partonic $\sqrt{\hat s}=10~\text{TeV}$ machines. Furthermore, the direct implementation of the sum rules does not assume an a priori large continuum contribution that often forms as a motivation for SMEFT interpretations.

In the following, we will consider a saturation of the sum rules with a minimal set of states in addition to the SM Higgs. This gives rise, as a function of the observed Higgs coupling modification $\kappa_H=g_{WWH_1}/(g_w m_W)$, to the two scenarios we study in this work: 
\begin{enumerate}[(i)]
\item the presence of vector states dominating the restoration of unitarity far above the Higgs resonance and 
\item the presence of additional scalar states to restore unitarity at large energies (Higgs mixing). In either case, the phenomenology is determined entirely by the resonance mass scale and $\kappa_H$, thereby creating a highly predictive framework that enables comparisons of high-energy colliders based on modelling probability conservation. In particular, the appearance of charged resonances $W'^\pm \to W^\pm Z$ leads to highly non-standard signatures in weak boson fusion that is ubiquitous at $\sqrt{\hat s}=10~\text{TeV}$ colliders.
\end{enumerate}
This list and our motivation are not exhausting all phenomenological possibilities, and analyses that are similar in spirit have been presented in, e.g.~\cite{Mohan:2015doa,Thamm:2015zwa,Liu:2018hum,Baker:2024xwh}.

Similar to the selection rules for unitarity in the gauge boson sector, the interactions of massive gauge bosons with fermions can lead to non-trivial constraints that correlate vector boson and Higgs couplings with massive fermions. In our analysis, for which initial states are at the parton level, the dominant effects are enhanced widths of heavy BSM particles. We will return to the impact of these potential effects on WBF analyses further below.

\section{Simulation and Analysis Setup}
\label{sec:sim}
\subsection{Benchmark scenarios}
\label{sec:benchmarks}
We investigate three different weak boson scattering signatures: $W^+W^-$ production, $W^\pm Z$ production, and $ZZ$ production, which arise as a consequence of the sum rules in Eq.~\eqref{eq:unitr} given an assumed deviation from the SM Higgs boson signal strength, i.e $\mu_H<1$. (In the following phenomenological analysis, we use the Higgs signal strength $\mu_H= \kappa_H^2$.)

Firstly, we consider a purely scalar sector extension to the SM with a heavy Higgs boson partner, $H^\prime$, which can be observed as a resonance in $VV\rightarrow W^+W^-$ and $VV\rightarrow ZZ$ channels. For a given choice of the $H^\prime$ mass and $\mu_H$, the coupling of the $H^\prime$ to the weak bosons is determined by unitarity requirements according to the sum rules. The $H'$ decay width $\Gamma_{H'}$ is a free parameter of the model that will ultimately determine the significance of the observation, but we choose values that ensure the $H^\prime\rightarrow VV$ decays dominate. We then scan the critical coupling required to satisfy the sum rules for specific choices of the $m_H^\prime$, $\Gamma_{H'}$ and $\mu_H$. Note that in this simplified Higgs-portal-like model~\cite{Binoth:1996au,Schabinger:2005ei,Patt:2006fw,Englert:2011yb} there is a coupling between the heavy Higgs boson and two SM-like Higgs bosons. However, unitarity favours the $H^\prime$ to show an enhanced coupling to (longitudinal) massive gauge bosons as a function of the $H^\prime$ mass, whereas perturbativity in the extended Higgs sector does not impose any such strong correlation. 

Secondly, we consider extending the SM with only iso-triplet states, $ V^\prime= W',Z'$, which can be observed as resonances in the $W^\pm Z \rightarrow W^\pm Z$ and $W^+W^-\rightarrow W^+W^-$ channels, respectively. Any deviation of $\mu_H$ from unity directly determines the $W'$ and $Z'$ couplings to SM gauge bosons and thereby their decay widths. Deviations from the SM-like $WWZ$ coupling are expected and have been calculated in holographic approaches~(e.g.~\cite{Englert:2008tn}), and they turn out to be parametrically small (in support of precision measurements). For practical purposes, we therefore choose the $WWZ$ coupling to be equal to the SM value and solve the sum rules for a given choice of $\mu_H$ and $m_{V^\prime}$

Nonetheless, this model remains an effective theory and unitarity is not guaranteed.\footnote{It is worthwhile highlighting that the scalar extension is not necessarily renormalisable; strongly interacting composite theories with additional light degrees of freedom can have similar phenomenological implications, see e.g. \cite{Gripaios:2009pe,Ferretti:2013kya,Appelquist:2022qgl}.} A phenomenologically relevant implication of this observation is the relation of the resonances with the fermion sector. In this second scenario, it should be noted that $W^+ W^- \to t\bar t$ scattering can lead to non-negligible couplings of the $Z'$ to SM fermions. Satisfying the sum rules imposed by vanilla $WW \to t \bar t$ scattering, including the new $WWZ'$ contribution and allowing $\mu_H<1$ can lead to a significant partial decay width for $Z'\to t\bar{t}$.\footnote{In an approach inspired by Eq.~\eqref{eq:unitr}, the strength of the $Z^\prime$ coupling to left-handed fermions is given by
\begin{equation*}
\mu_{Z',L} = {3 c_W \over (4 c_W^2 -1) s_W} {\kappa_H -1\over g_{WWZ'}} 
\end{equation*}
when expressed in units of the $Z$ boson coupling in the SM. The coupling to right-handed fermions is zero. Here, again, $\kappa_H^2=\mu_H$, and $g_{WWZ'}$ is fixed by the gauge boson scattering sum rules.} This decay mode can, in principle, outpace the $Z'\to WW$ decay mode, but the extent to which that is true depends on the details of the UV-complete model, see, e.g.~\cite{Ferretti:2014qta}. The results presented in this paper for $W^+ W^- \to Z^\prime \to W^+ W^-$ could therefore weaken, but would be compensated by new $Z'\to t\bar{t}$ discovery modes.

\subsection{Monte Carlo simulation and analysis strategy}
\label{sec:ana}
We generate events using \texttt{MadGraph5\_aMC@NLO}~\cite{Alwall:2014hca}. The production of $W^+W^-$, $W^\pm Z$ and $ZZ$ via weak boson scattering is calculated using a custom model implemented in \texttt{FeynRules}~\cite{Alloul:2013bka} and with a \texttt{UFO}~\cite{Degrande:2011ua} interface. The results are cross-validated against~\cite{Arnold:2008rz}. Our model allows weak boson scattering predictions to be obtained at both muon colliders and hadron colliders for the different benchmark scenarios discussed in Sec.~\ref{sec:benchmarks}. The model also predicts the non-resonant (SM) weak-boson scattering process as well as the interference between the resonant and non-resonant contributions.

\begin{figure*}[!t]
\centering
\subfigure[]{\includegraphics[height=0.34\textwidth]{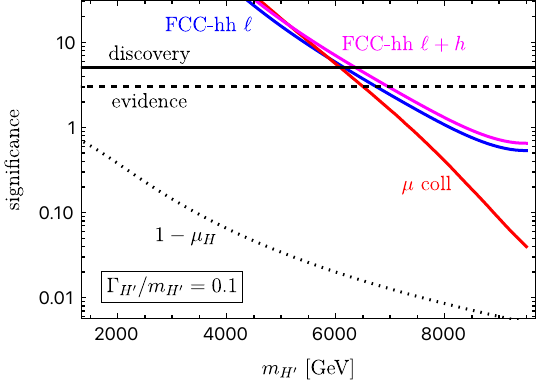}}
\hfill
\subfigure[]{\includegraphics[height=0.34\textwidth]{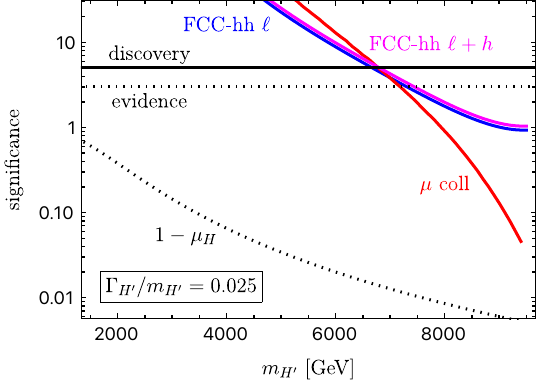}}
\caption{\label{fig:hexo} Sensitivity of resonance searches in the $WW$ and $ZZ$ weak boson fusion production modes when considering scalar unitarity restoration for $\mu_H<1$. We show two scenarios for comparison related to the width of the (assumed narrow) resonance (a) $\Gamma_{H'}/m_{H'}=10\%$ and (b)  $\Gamma_{H'}/m_{H'}=2.5\%$; overlaid on the significance plot is the signal strength that such parameter choices will probe. The magenta lines include the expected improvements from accessing semi-leptonic channels (FCC-hh efficiencies are provided at the end of Sec.~\ref{sec:ana}, we comment on more conservative current LHC efficiencies in the text).}
\end{figure*}

At hadron colliders, the experimental signature of weak boson scattering (WBS) is the production of two weak bosons in association with two jets that are separated by a large rapidity interval. Our proposed resonance-search strategy is to tag the two jets and exploit the leptonic decays of the weak bosons, i.e. similar to current WBS measurements at the LHC~\cite{ATLAS:2018mxa,ATLAS:2019cbr,ATLAS:2020nlt}. There are two important sources of background: (i) non-resonant SM WBS production of $VVjj$ and (ii) $VVjj$ production at order $\alpha_S^2\alpha_{EW}^2$ (typically mediated by the $t$-channel exchange of a gluon). We simulate both of these processes directly to estimate the total background.  

The LHC experiments have also shown sensitivity to weak boson scattering in semi-leptonic final states when using $V\rightarrow jj$ tagging algorithms~\cite{ATLAS:2025omi}. In these final states, the hadronic weak boson decay can be fully reconstructed, and this aids in the resonance identification, so we include this channel in our analysis. The backgrounds from non-resonant SM WBS production of $VVjj$ and $VVjj$ production at order $\alpha_S^2\alpha_{EW}^2$ are again simulated and included in the analysis. However, for semi-leptonic final states, $V+\textrm{jets}$ production acts as an important additional background because two of the jets can be misidentified as a $V\rightarrow jj$ candidate. We do not simulate this directly, but instead scale the non-resonant SM WBS background yield by a factor $R_{\textrm{bkd}}$ estimated from recent ATLAS analyses.\footnote{The values of $R_{\textrm{bkd}}$ are taken to be $R_{\textrm{bkd}}=6$ and $R_{\textrm{bkd}}=20$ for two-lepton and one-lepton final states, respectively. $R_{\textrm{bkd}}$ is estimated using the ratio of the $V+\textrm{jets}$ event yield to the WBS event yield that was observed in a recent ATLAS semi-leptonic analysis~\cite{ATLAS:2025omi}, whilst also accounting for performance improvements in $V\rightarrow jj$ tagging algorithms that have recently been obtained using transformer architectures~\cite{ATLAS:2023zcb}.} The purely hadronic decay channels are swamped by huge multijet backgrounds, and we do not consider them further in our analysis.

At a muon collider, multijet-induced backgrounds only arise from $VV\rightarrow Vjj$ and $VV\rightarrow 4j$ production and are therefore heavily suppressed when compared to a hadron collider. We therefore study inclusive final states at the muon collider and profit from (i) much larger branching ratios and (ii) the fully reconstructed final state for $V\rightarrow jj$ decays. We have simulated the backgrounds from $VV\rightarrow Vjj$ and $VV\rightarrow 4j$ production and found that they are negligible compared to the non-resonant SM WBS process. We note that top-antitop production can be suppressed with a combination of $b$ vetoes and the identification of the top-quark from its fully-hadronic final state.

\subsection{Event selections and sensitivity estimates}
The weak bosons from the resonance decays are produced centrally with very high transverse momentum and therefore easily pass typical event and trigger selections, such as $p_{T,\ell} > 20$~GeV and $|y_\ell|<2.5$ for charged leptons, $p_{T}^{\text{miss}}>50$~GeV for final states containing neutrinos, and $p_{T}^{\text{miss}}>400$~GeV for $V\rightarrow jj$ decays. For hadronic weak boson decays, we assume a 50\% $V\rightarrow jj$ tagging efficiency at hadron colliders~\cite{ATLAS:2023zcb} and an 80\% tagging efficiency at the muon collider. Here we have assumed that a looser working point can be adopted at a muon collider due to the much smaller continuum multijet backgrounds. Finally, at the hadron collider, events are also required to have a WBF-like topology, i.e. two tagging jets with $p_{T,j} > 30$~GeV and $|y_j|<4.5$ that also satisfy $m_{jj}> 900$~GeV and $\Delta y_{jj}>4$. At the muon collider, WBF is the dominant production mechanism, and no additional selections are imposed. 

We exploit two discriminating variables in our resonance: (i) the diboson invariant mass for fully reconstructed final states and (ii) the diboson generalised transverse mass for final states containing neutrinos. The generalised transverse mass is defined as
\renewcommand{\vec}{\bf}
\begin{equation}
m^{2}_{T}= \left[ \sqrt{m^{2}_{\textrm{vis}}+p_{T, \textrm{vis}}^{2}} +  |p_{T}^{\text{miss}}| \right]^2 - \left[ {\vec{p}}_{\textrm{T}, \textrm{vis}}^{} +{\vec{p}^{}}_{T}^{\text{miss}} \right]^2\, ,
\end{equation}
where `vis' implies the four-vector sum over all visible decay products from the two weak bosons (i.e. charged leptons and jets). 

For $WW\rightarrow \ell\nu\ell\nu$, there is significant missing energy from the two neutrinos in the final state, and this leads to a broad Jacobian peak in the transverse mass spectrum. We only investigate this final state at the hadron collider, where we select events with
\begin{equation}
m_T\in [0.5,2]~m_{H'/Z'}\,.
\end{equation}
For $WV\rightarrow \ell\nu jj$ and $WZ\rightarrow \ell\nu\ell\ell$, there is only one neutrino in the final state, and the transverse mass distribution exhibits a much narrower Jacobian peak. We select events with
\begin{equation}
m_T \in [m_{V'} - 5 \Gamma_{V'} ,  m_{V'} + 2 \Gamma_{V'} ]\,.
\end{equation}
These final states are experimentally accessible at both hadron and muon colliders. For $VV\rightarrow 4j$, $VZ\rightarrow jj\ell\ell$ and $ZZ\rightarrow 4\ell$ production, the fully reconstructed final state allows the use of the diboson invariant mass directly. Events are selected with 
\begin{equation}
m_{VV} \in [m_{X'} -2 \Gamma_{X'}, m_{X'}+2\Gamma_{X'}]\,,
\end{equation}
where $X^\prime = H^\prime,W^\prime$. The $VV\rightarrow 4j$ final state is only considered at the muon collider.

\begin{figure*}[!t]
\centering
\subfigure[\label{fig:zp}]{\includegraphics[height=0.34\textwidth]{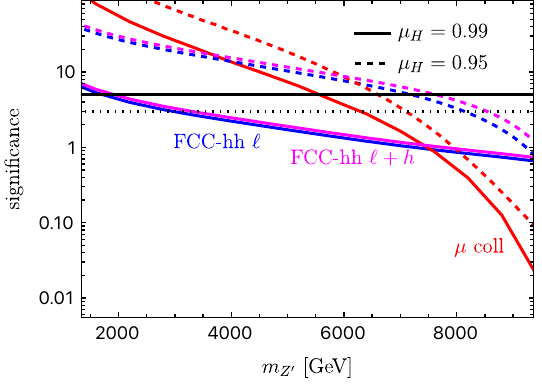}}\hfill
\subfigure[\label{fig:wp}]{\includegraphics[height=0.34\textwidth]{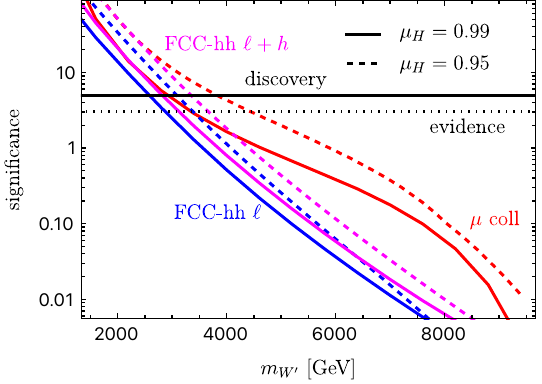}} 
\caption{Sensitivity of a WBF search of unitarity-restoring vector resonances. Shown are (a) the $Z'$ search and (b) the $W'$ search for two choices of Higgs signal strengths $\mu_H=0.95,0.99$. A significant coupling modification leads to a large signal at an FCC-hh, which decreases more quickly as it approaches the SM, since the kinematic endpoint of the FCC-hh is probed with a lower statistical yield compared to the muon collider. The magenta lines include the expected improvements from accessing semi-leptonic channels.\label{fig:vp}}
\end{figure*}

\begin{figure}[!t]
\centering
\vspace{-0.3cm}
\includegraphics[height=0.34\textwidth]{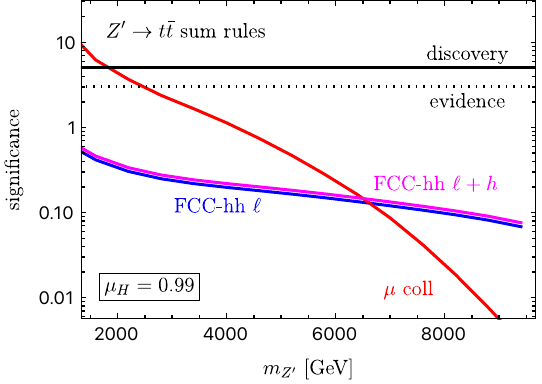}
\caption{\label{fig:zpwide} Sensitivity for wide fermion-aware resonances. The magenta line includes semi-leptonic channels using the FCC-hh extrapolation of efficiencies given in the text.}
\vspace{-0.2cm}
\end{figure}

Throughout, we consider a muon collider operating at $\sqrt{s}=10$~TeV with a target luminosity of $10~\text{ab}^{-1}$ and FCC-hh operating at $\sqrt{s}=100~\text{TeV}$ with a total integrated luminosity of $30~\text{ab}^{-1}$.  In all cases, we estimate the significance at a given machine and luminosity using ${S/\sqrt{B}}$ where $S$ is the number of signal counts obtained from the selection signal cross section and the luminosity, and $B$ is the background count in the same region. We have checked that the event yields are sufficient for this approximation to hold for the claimed discovery regions. Significances of independent channels are treated as uncorrelated. `Evidence' and `discovery' refer to significances of $3\sigma$ and $5\sigma$, respectively. 

Systematic uncertainties have not been accounted for in our significance estimates in detail. For a resonance search, the backgrounds can likely be constrained using fits to the $m_{VV}$ (or $m_T$) spectrum, either side of the resonance peak. This would likely be done in a profile-likelihood approach, which would also allow the signal shape to be exploited. At the FCC-hh, control regions would be used to constrain the modelling of non-WBS backgrounds. At this stage, many decades before the FCC-hh or a muon collider would be operational, it is not possible to accurately estimate the modelling systematics. Instead, our results should be understood as being indicative of the relative performance of FCC-hh and a muon collider when searching for resonances associated with unitarity restoration. 

\section{Results}
\label{sec:res}
Starting with exotic Higgs production, we show results in Fig.~\ref{fig:hexo} for two scenarios: $\Gamma_{H'}/m_{H'}=0.025 \,\, \textrm{or} \,\, 0.1$. Both FCC-hh and the muon collider have very similar sensitivity and are capable of observing scalar resonances with a mass of up to 6~TeV. Unitarity forces the coupling from Higgs mixing to $g_{VVH'}\simeq 8\times 10^{-3} g_{VVH}^\text{SM}$. However, the origin of the sensitivity is very different at each collider. Specifically, $H'\to WW$ drives the sensitivity, and although the hadron collider has a much larger overall cross section for resonance production via weak boson fusion, the fully-leptonic signal is smeared over a large window in transverse mass due to the presence of two neutrinos in the final state. This leads to substantial background contamination at FCC-hh. The inclusion of the semi-leptonic decay channel at FCC-hh does not improve the sensitivity, despite the narrower Jacobian peak in the transverse mass spectrum, due to the very large background from $V+\textrm{jets}$ events. The muon collider, on the other hand, exploits the fully hadronic decays of the weak bosons to reconstruct the diboson invariant mass directly, allowing the narrow resonance to be identified with much less background contamination. It is interesting to note that FCC-hh and muon collider sensitivities are comparable in the parameter region of narrow resonance searches. This is due to efficient background rejection at the hadron machine and efficient operation at a signal-dominant muon collider. Broad resonances are not covered in this work; they lend themselves to continuum-enhanced searches building on EFT analyses. However, it is worthwhile mentioning that for increasingly broader resonances, one needs to rely on a larger $m_{VV}$ window at the muon collider, which leads to a substantial decrease in sensitivity. The FCC-hh sensitivity is almost unaffected, however, because the transverse mass window is driven by the inability to reconstruct the final-state neutrinos. 

The sensitivity to vector resonances is shown in Fig.~\ref{fig:vp}. The hadron collider and muon collider would have similar sensitivity to a $Z^\prime$ resonance if large coupling deviations are observed at the HL-LHC ($\mu_H=0.95$), probing resonance masses up to roughly 7~TeV (this corresponds to a coupling $g_{WWZ'}\simeq 1.0\times 10^{-3}$). For small coupling deviations ($\mu_H=0.99$), the muon collider would retain sensitivity up to about 5.5~TeV ($g_{WWZ'}\simeq 0.5\times 10^{-3}$), whereas the FCC-hh has no sensitivity above resonance masses of about 2~TeV ($g_{WWZ'}\simeq 1.7\times 10^{-3}$). The sensitivity to a $W^\prime$ resonance is similar at both colliders, with sensitivities around 2.5-3.0~TeV for $\mu_H=0.99$ corresponding to $g_{W'WZ}\simeq 1.4\times 10^{-3} $, rising to up to roughly 4~TeV if $\mu_H=0.95$($g_{W'WZ}\simeq 2.4\times 10^{-3} $). This is predominantly due to a larger production cross section for larger values of $1-\mu_H$ that still lead to a relatively narrow decay phenomenology across the different decay modes. 

As mentioned earlier, we do not include a detailed analysis of the systematics that can arise in either collider environment. The discovery threshold selects within our assumptions a signal-rich region, which in turn makes our results relatively robust against flat background systematic effects. Irrespective of the model and collider we consider, including for example a flat 20\% variation of the expected background normalisation, leads to a modification of the discovery potential of the new states by 150 to 200 GeV.

We finally return to the theoretical considerations related to fermion-gauge-Higgs unitarity. The $Z'$ discovery reach in the considered WBF production mode is crucially sensitive to couplings to fermions. If the extra vector resonances are fermiophobic, both muon and hadron colliders show formidable sensitivity as shown in Fig.~\ref{fig:vp}(a). If there is a significant coupling of these states to fermions, the additional contributions to the decay width from $Z'\to t \bar{t}$ suppress the overall rate of $Z'\to W^+ W^-$ and restrict the sensitivity at both colliders dramatically, as shown in Fig.~\ref{fig:zpwide}. A muon collider in the described analysis setup would still retain some sensitivity. In this case, however, attention would shift to analyses of $t\bar t$ production, not covered in this work.

\section{Conclusions}
\label{sec:conc}
In this work, we have highlighted how a two-pronged approach combining Higgs precision measurements with high-energy searches for iso-scalar and iso-vector resonances (expected in theories of strong electroweak symmetry breaking) has the potential for new physics discovery at this high-energy frontier. Our analysis is solely based on unitarity and convergence, leading to a potential no-lose theorem at FCC-hh and a 10 TeV muon facility in case a small coupling deviation of the Higgs from the SM expectation is realised in nature (and potentially observed at the HL-LHC). In such an instance, narrow resonances could show up in the multi-TeV region, and searches for these particles are complementary to EFT-based approaches that are applicable to wide resonances at energies comparable to the maximally attainable collider energy scale. We have aimed to provide a qualitative picture of the sensitivity achievable at these experiments within the context of bump hunting in WBF production channels (which remove associated production modes not considered in this work). Of course, many of the experimental details of these future colliders are currently not known. Keeping these limitations in mind, we find that the physics potential for observing unitarity-restoring resonances is excellent at both FCC-hh and at a 10 TeV muon collider. The next community-agreed steps will likely involve an FCC-ee with dedicated precision $Z$ pole runs and a detailed investigation of $e^+e^- \to HZ$ production. This machine can therefore probe the $HVV$ vertices with unparalleled precision~\cite{Banerjee:2019pks,Banerjee:2021huv,FCC:2025lpp}. A dedicated, relatively model-independent measurement of the $HZ$ production could detect a $0.3\%$ departure from the SM~\cite{Banerjee:2019twi,Selvaggi:2025kmd}. This way, the FCC-ee can corroborate the findings of the HL-LHC, or can form the basis of a resonance search detailed in this work if the HL-LHC data set does not reveal a significant Higgs coupling deviation. Furthermore, the enhanced sensitivity that can be achieved through measurements of electroweak precision data can further reveal the presence of new interactions. For instance, a naive extrapolation of the oblique correction constraints of \cite{Baak:2014ora} extrapolated to a TeraZ machine shows sensitivity to Higgs coupling modifiers of ${\cal{O}}(10^{-3})$ in the renormalisable scalar mixing extension, where these constraints are particularly reliable. This demonstrates that a significant deviation observed at FCC-ee will therefore make a strong case for future high-energy facilities. Further related phenomenological analyses have been provided for composite theories in~\cite{Fabbrichesi:2007ad,Falkowski:2007iv,Cheung:2008zh,Barbieri:2008cc,CarcamoHernandez:2010qxf,Bellazzini:2012tv,Chang:2013aya,Espriu:2014jya,Fleper:2016frz,BuarqueFranzosi:2017prc}.

\medskip
\noindent {\bf{Acknowledgments}} ---
C.E. is supported by the Institute for Particle Physics Phenomenology Associateship Scheme. 
W.N. acknowledges support by the Deutsche Forschungsgemeinschaft (DFG, German Research Foundation) under Germany’s Excellence Strategy - EXC 2121 ``Quantum Universe'' - 390833306. This work has been partially funded by the Deutsche Forschungsgemeinschaft (DFG, German Research Foundation) - 491245950.
A.D.P. is supported by the STFC under grant ST/W000601/1 and by the Leverhulme Trust under grant RPG-2020-004.
M.S. is supported by the STFC under grant ST/P001246/1.
\bibliography{references}
\end{document}